\documentclass[12pt,a4paper]{article}

\usepackage{graphics}
\usepackage{latexsym}
\usepackage{amsmath}
\usepackage{amssymb}

\title{Built to evolve}
\author{
B. Hoeneisen and G. Trueba
}
\date{\small{Universidad San Francisco de Quito \\
15 June 2005} }

\begin{document}
\maketitle

\begin{abstract}
\noindent
We study the probabilities of evolution based on 
random mutations and natural selection. We 
conclude that evolution
to multicellular eukaryots, or even prokaryots, is 
unlikely to be the result of only
random mutations. Complex
organisms have evolved through several mechanisms 
besides random mutations, namely DNA recombination, 
adaptive mutations, and acquisition of foreign DNA.
We conclude that all living organisms, 
in addition to being self-organizing and reproducing
(autopoyetic), have built-in mechanisms of evolution, some of which 
respond in very specific ways to environmental stress.
\end{abstract}

\section{Introduction}
We consider the probabilities of obtaining genomes by random mutations
and natural selection, and obtain the conditions that are required
for successfull evolution. Examples are presented for 
viruses, bacteria, eukaryote cells, and multi-cellular organisms.
Our conclusions are collected in Section 10.

\section{The odds}
The human genome has about 30 thousand useful genes, each with 
an average coding region of 20 thousand base pairs coding about 6666
aminoacids.\cite{nature} What is the probability of writing
a specific sequence of 
$30000 \times 6666$ words (aminoacids) chosen at random from a list of 20?
The answer is $20^{-30000 \times 6666} \approx 10^{-260000000}$, i.e. zero for
all practical purposes. 
For that matter, what is the probability
of getting the correct genome of any one of the 30 million species,
with one try every second since the Big Bang,
by every one of the $10^{31}$ (or so) bacteria on Earth?
The answer is 
$3 \times 10^7 \times 4 \cdot 10^{17} \times 10^{31} \times 20^{-30000 \times 6666}$
\textbf{which is still the same result}: $\approx 10^{-260000000}$
(only the last two digits in the exponent are changed), i.e. zero for
all practical purposes.

Note that the \textquotedblleft{simplest}" living organism, the prokaryote bacteria,
is almost as complex as a human being: it has 
of order 1000 active genes with an average of $\approx 1500$ base pairs!
There are many missing links between organic molecules (sugars, lipids, 
bases, aminoacids, etc) and the simplest forms of life. 

Perhaps the trick is to write little pieces of genetic code at a time. 
So, let us turn the question around: What is the largest gene that
can be produced at random with a finite probability (say, $10^{-4}$) with
one try every second during 100 million years by each of 
the $10^{31}$ bacteria on Earth?
The answer is one gene encoding, at most, $\approx 39$ aminoacids,
or $\approx 117$ base pairs.
This is roughly the limit for undirected random evolution.

\section{Evolution}
Let us play a game called \textquotedblleft{Evolution}".
We sit in front of a key board and hit keys at random. 
The probability of obtaining \textquotedblleft{Romeo and
Juliet}" is zero for all practical purposes. In fact,
the probability of obtaining any meaningfull novel in any
known language is zero for all practical purposes.
Now introduce \textquotedblleft{mutations}", i.e. 
replace random letters by random hits of the keyboard.
Still no meaningfull novel will ever be obtained
for all practical purposes. 

Now suppose we are allowed to select 
which letter (or small set of letters) to mutate at a rate much
higher than the background mutation rate of the other letters,
and we are allowed to stop the hypermutations when a particular 
outcome is obtained. For example, choose to hypermutate
the 10th letter until \textquotedblleft{e}" is obtained,
or choose to hypermutate the 10th, 11th and 12th letters
until \textquotedblleft{dog}" is obtained.
Of course, now we can write any novel at all: the
game has become trivial. If the choice of which letter
to hypermutate is perfectly specific, and the choice
of which outcome to select is perfectly specific, the game
becomes trivial, i.e. the outcome becomes certain,
even tho the keys are hit at random.
If the environment were perfectly specific, it would have
perfect control over evolution (even if mutations occur
at random). 

Evolution lies somewhere in between. The choice
of which bases on which genes to hypermutate is not perfectly 
specific and is incomplete, and the outcome that stops the
hypermutations is also not perfectly specific. However,
with enough specificity it is possible to write little 
pieces of survivable genome. Then the pieces can be combined.
In our example we could copy \textquotedblleft{dog}",
reverse \textquotedblleft{dog}",
concatenate \textquotedblleft{dog}" and \textquotedblleft{cat}",
interchange \textquotedblleft{dog}" and \textquotedblleft{cat}"
(as in horizontal gene transfer, or in meiosis), 
introduce words and phrases from other books, and so on.
The interchange of genetic
material within, and between, bacteria, viruses and eukaryote cells plays
a major role in evolutionary change.\cite{virus} 

Survivable pieces of the genome will often involve negative
(stabilizing) feedback loops. A hierarchy of negative 
feedback loops, within negative feedback loops, within ...,
is self organizing. 

\begin{figure}
\begin{center}
\vspace*{-1.2cm}
\scalebox{0.5}
{\includegraphics{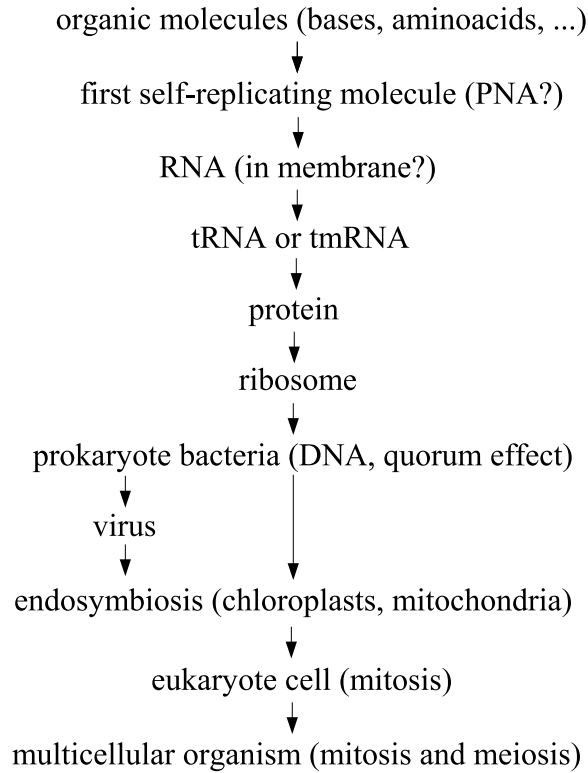}}
\vspace*{-0.8cm}
\caption{Plausible steps of evolution.
For PNA see \cite{RNA}, for tRNA see \cite{tRNA, tmRNA},
for genetic recombination see \cite {recombination}, for viruses see
\cite{virus}.}
\label{ev}
\end{center}
\end{figure}

The steps in evolution, from simple to complex, might have been as shown in 
Figure \ref{ev}. At all levels of complexity the environment
must have directed evolution with sufficient specificity for the steps
to have a non-zero probability.

\begin{figure}
\begin{center}
\vspace*{-0.3cm}
\scalebox{0.6}
{\includegraphics{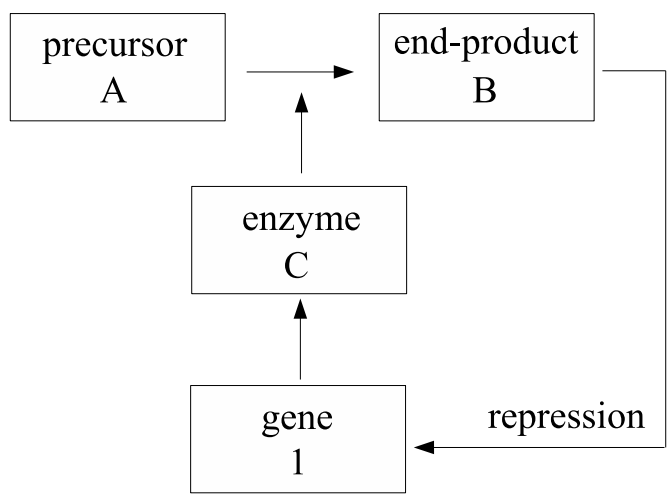}}
\vspace*{-0.3cm}
\caption{Precursor A is converted to end-product B by enzyme C.
Enzyme C is encoded by gene 1.
The concentration of B is regulated by
the repression of gene 1 by the end-product B.
Note that enzyme C is synthesized only when needed.
This circuit is regulating and is built to evolve.}
\label{derepression}
\end{center}
\end{figure}

\section{Adaptive mutations}
Consider the (simplified) metabolic pathway of a cell
shown in Figure \ref{derepression}.
Precursor A is converted to end-product B by enzyme C.
Enzyme C is encoded by gene 1. The transcription of gene 1 
(and synthesis of enzyme C) is repressed (directly or indirectly)
by the end-product B. This negative feedback loop regulates the
concentration of B, and is efficient because enzyme
C is produced only when needed. This feedback loop 
therefore has a value for
survival, and would have been selected by nature. Now starve the cell
of precursor A in the presence of a similar precursor A$'$. 
The result is a reduction of the concentration
of the end-product B, and a derepression of gene 1. 
The rate of transcription of gene 1
increases (by a factor that can exceed 1000\cite{Wright}).
During transcription, mRNA copies one strand of DNA exposing the other
strand. Single strand DNA is prone to mutations due to the lack
of hydrogen bond stabilization between complementary bases,
the formation of loops and other secondary structures,
\footnote{Segments of the single strand DNA may stick to other segments
with mostly complementary bases,
resulting in unpaired or mispaired
bases. Unpaired bases are prone to deamination, deletion or
replacement. Cytosine deaminates to uracil at a rate 100 times larger
in single strand DNA than in double strand DNA.\cite{Wright}
Mutations also occur in the end-loops where bases have no
complement, and in the stem.
These errors are immortalized during DNA duplication or repair.}
and supercoiling.\cite{Wright} 
As a result, gene 1
acquires a high rate of mutations and begins synthesizing enzymes
similar to C, until one of them is able to
convert precursor A$'$ into end-product B. Gene 1 is then repressed by
end-product B and hypermutation stops. 
The resulting negative feedback loop resumes control
of the concentration of the end-product B to the same original
concentration. The net result
of these processes is that a change of the environment
(the starvation for precursor A) triggers mutations
of a specific gene of the cell, until the cell 
is capable of substituting precursor A by precursor A$'$.
So the environment can direct evolution in very specific ways.

Let us briefly describe examples.

\section{B-lymphocytes}
\begin{figure}
\begin{center}
\vspace*{-1.0cm}
\scalebox{0.5}
{\includegraphics{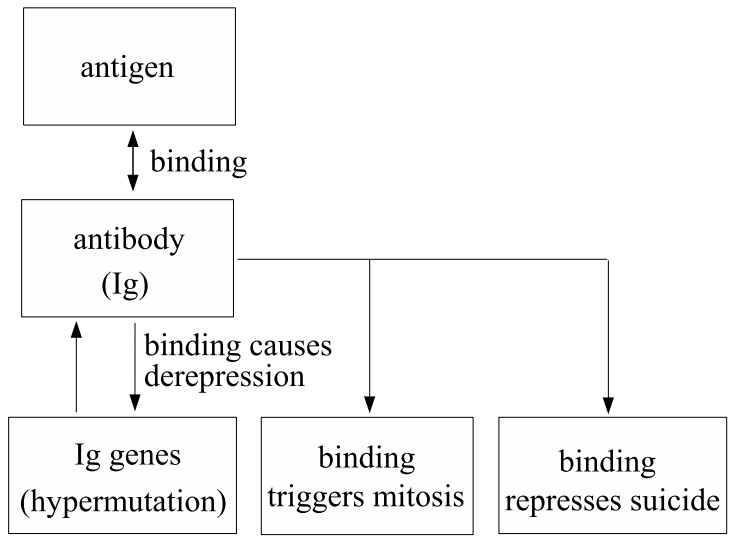}}
\vspace*{-0.4cm}
\caption{B-lymphocytes have antibody proteins (called
immunoglobulins, Ig) attached to their
membrane. Binding of these proteins to specific antigens
triggers mitosis of the B-cell. Hypermutation of the
Ig coding genes causes variation in the antibodies.
Binding to the antigens rescues the B-cell from programmed
death. The result is survival of the B-cells producing the
most specific antibody.}
\label{B_cell}
\end{center}
\end{figure}

Let us consider B-lymphocytes of the
immune system (see Figure \ref{B_cell}),
which have been studied in considerable detail.\cite{Immunology, hypermutation}
These B-lymphocytes have antibody
proteins (called immunoglobulins) attached to their 
membrane. An invading bacteria has antigen proteins
attached to its membrane. The antibody can bind
to a very specific set of antigens. This binding triggers 
a series of complex steps (including helper
T-lymphocytes) that activate mitosis of the B-cell,
expresses several genes that code immunoglobulins,
and differentiates the B-lymphocytes into antibody secreting
cells and memory cells.
Proliferanting B-cells in germinal centers show high
rates of point mutations in genes coding immunoglobulins
($10^3$ to $10^6$ times higher than the spontaneous rate 
of other genes).\cite{hypermutation} The result is the synthesis of
immunoglobulins with small differences. The binding of
these antibodies to the antigens (with the intervention
of follicular dendritic cells) produce signals that
rescue the B-lymphocytes from programmed cell death.
At the latter stages of the infection the concentration of
invading bacteria becomes low, so only B-cells
producing very high affinity antibodies can bind to the
antigens and survive. This phenomenon is called
\textquotedblleft{affinity maturation}".
 
The net result of these complex processes is that a change of
the environment (the invading bacteria) triggers
hypermutations  of specific sections (those that code for
the binding site of the immunoglobulins) of specific genes of specific
B-cells, and selects mutations producing immunoglobulins with
the highest affinity to the antigen. Note that each one of these
steps is very specific.

Since B-lymphocytes are somatic, the selected mutations
are not passed on to the next generation, so this 
is an example of evolution of B-cells in one individual, not
evolution of the species. 

The hypermutation associated to affinity maturation of B-cells
is caused by the induction of mutagenic genes such as cytidine
deaminase (which causes C to U transitions) and 
error prone DNA polymerases. The presence of the same genes
in other eukaryots may indicate that similar hypermutagenic
processes may occur in eukarya, and, as in prokaryots or B-cells,
the induction of these genes may be triggered by environmental stress
such as starvation.

Let us mention that B-lymphocytes with hypermutation
and recombination of V, D and J cassettes
of genes coding immunoglobulins, produce
B-cells that synthesize of order $10^{11}$ different 
antibodies capable of binding to as many different
antigens. So, with just a handfull of genes in the genome,
B-cells are able to synthesize a much larger number 
of different proteins!

%How are lymphocytes able to tell self-proteins from
%foreign proteins? It turns out that at an early age of the 
%fetus, binding causes suicide of the lymphocytes (instead
%of preventing suicide), until no
%lymphocytes remain that bind to self-proteins.

\section{\textit{Enterobacter arogenes}}
\begin{figure}
\begin{center}
\vspace*{-1.0cm}
\scalebox{0.5}
{\includegraphics{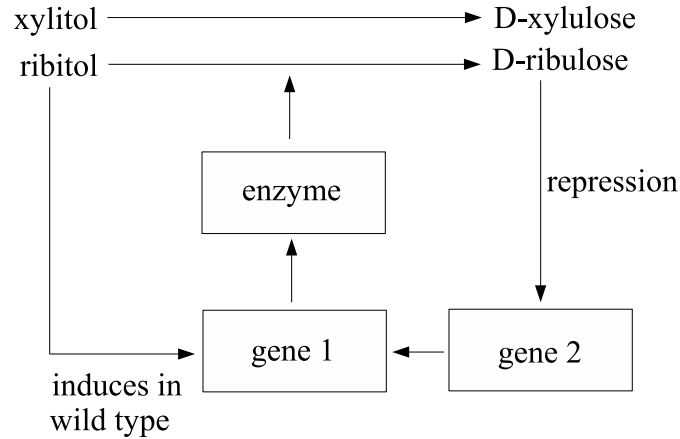}}
%\vspace*{-0.2cm}
\caption{In the wild strain 5P14 of
\textit{Enterobacter arogenes}
ribitol induces gene 1 to synthesize  
ribitol dehydrogenase. This enzyme metabolizes
ribitol, and, with low specific activity, xylitol.
Starvation for ribitol leads to
a mutation of gene 2 which causes the expression of
gene 1 even in the absence of ribitol. Hypermutations
of gene 1 and natural selection result in a more specific enzyme.}
\label{enterobacter}
\end{center}
\end{figure}

Let us briefly describe the metabolic pathway of
\textit{Enterobacter arogenes} shown in
Figure \ref{enterobacter}.\cite{Lerner, Wu, Wright}
In the wild strain 5P14,
ribitol induces gene 1 to synthesize
ribitol dehydrogenase. This enzyme metabolizes
ribitol, or, with low specific activity, xylitol.
The wild bacteria can therefore metabolize xylitol only
if ribitol is present. If the bacteria is starved for
ribitol in the presence of xylitol, a mutation occurs
in a gene 2 that causes the expression of gene 1 even in the absence of
ribitol. This strain, called X1, appears in 4.1 hours.
Mutations in gene 1 produce strains X2 in 1.7 hours,
and then X3 in 0.9 hours. These strains synthesize
modified enzymes with increasing specific activity on
xylitol.

So a change of the environment (the removal of ribitol in the
presence of xylitol) results in specific mutations of two 
specific genes. The specificity is so great that the experiment
is repeatable!

\section{Sex}
Mate the largest dogs of different litters for several generations, and you
end up with huge Great Danes. This is artificial selection. A Great Dane
has many \textquotedblleft{big-dog-genes}". These big-dog-genes were already
in the genetic pool of the population of dogs.
The largest dog of a litter probably has more big-dog-genes than each 
parent due to the mixing of genes during sexual reproduction (meiosis). 
The smallest dog of
the litter probably has less big-dog-genes than each of the parents.

Natural selection can work in a similar way. An ecological niche attracts
individuals specially adapted to that niche. 
\footnote{Mimetism is an example: a green insect that chooses a green
environment to hide in has a better chance of survival.}
These individuals have an
enhanced probability of mating with each other. If the population has
a gene pool with several genes that favor the niche, then, after a 
few generations, these genes can come together and we obtain individuals
specially adapted to the niche. If these individuals no longer mate
and reproduce with
the general population, then a sub-species has formed.

Even cultural preferences can bias mating, resulting in genomes specially
adapted to these cultural preferences. This is known as the Baldwin
effect, after the description of this phenomenon by James Mark Baldwin in 
1896.

\section{Viruses}
A virus is composed of genetic material packaged (mostly) in
proteins. The genetic material may be linear or circular, 
single or double stranded, haploid or diploid, monopartite or
multipartite, DNA and/or RNA. The RNA can be
\textquotedblleft{positive}" and serve as a messenger RNA
to directly synthesize proteins (using the tRNA and ribosomes
of the host cell), or it can be \textquotedblleft{negative}"
and require a transcription to +RNA before protein synthesis.

What proteins are coded by the DNA or RNA?
In order to reproduce, the virus must code the
proteins that form part of the virus itself: structural proteins
and enzymes needed prior to protein synthesis
(such as enzymes used by retroviruses to synthesize DNA from RNA, enzymes used
by negative strand RNA viruses to transcribe -RNA to +RNA, 
enzymes used by double stranded RNA viruses to make single strands, etc).
Depending on the type of virus, other proteins may be coded as well.
Examples are enzymes to transcribe 
+RNA to -RNA, DNA to RNA at various starting sites, 
proteins that block defenses of the
host cell, proteins that cleave other proteins at special sites
(so one mRNA of the virus can code many proteins linked together
at cleaving sites). 

In addition, and of particular interest for
evolution, the virus may encode proteins that can turn on or turn
off hypermutations, and enzymes used to recombine RNA within
the virus, among different viruses (even of different species),
and between the virus and the host cell.
Let us quote from \cite{virus}:
\textquotedblleft{The two major forces acting upon viral genomes to generate
diversity that can be tested for environmental survival
and replicative fitness are mutations and recombination.
Some viruses have a good deal of control over their own rates of mutation and 
even the frequency of recombination. They exert control by encoding
viral enzymes" for replicative and recombinational functions.

\section{Other examples}
A strain (known as FC40) of \textit{Escherichia coli} can not digest
lactose due to a frameshift mutation in gene
\textit{lacZ} that does not allow the
synthetis of $\beta$-galactosidase 
in sufficient quantity.
It has been observed that this frameshift mutation
undergoes reversion when lactose becomes the sole source
of energy. It is important to note that most 
reversions occur after exposure to 
lactose.\cite{escherichia_coli}

Starvation of \textit{Escherichia coli} induces the production
of alternative polymerase enzymes (DinB and UmuD$'_2$C) which
are capable of replicating badly damaged sequences of DNA,
and, in the process, produce high rates of 
mutations.
Homologs of these alternative enzymes have been found in
\textit{Saccharomyces cerevisiae}, mice and 
humans.\cite{adaptive_evolution}

\textit{Escherichia coli} is able to mutate even when not
dividing or replicating its DNA, and these mutations may be its main
source of genetic variation.\cite{escherichia_coli}

Some plants switch from asexual proliferation (rhizomes) to 
sexual reproduction in conditions of stress. In doing so they
speed up evolution
by trying new combinations of genetic material 
in the process of meiosis.

Snails switch from hermaphrodite reproduction to bi-sexual
reproduction in conditions of stress due to parasites.
This strategy speeds up evolution when needed.

\section{Conclusions}
Evolution appears to be hopelessly improbable
unless random mutations are limited to no more than about 100 bases
of specific genes, and the selection of the outcomes are
sufficiently specific. We therefore propose that all living
organisms, in addition to being self-organizing 
and reproducing (autopoyetic),
are built to evolve in selective ways.
It appears that viruses, prokaryots and eukaryots have
considerable control over the rates and spectrum of mutations
and recombinations.
There are built-in mechanisms to control hypermutations of
selective regions of selective genes, and sufficiently selective
mechanisms to choose the outcome of these mutations. The
high selectivity of these and other mechanisms are required for evolution
to be successfull.
We have given examples of viruses, prokaryots, eukaryots, and multicellular
organisms, where this is indeed the case.

\end{document}